\documentclass[structabstract]{aa}
\usepackage{graphicx}
\usepackage{txfonts}
\usepackage{natbib}
\usepackage{hyperref}
\usepackage{color}

\begin{document}
   \title{Modeling optical and UV polarization of AGNs}

   \subtitle{V. Dilution by interstellar polarization and the host galaxy}

   \titlerunning{Modeling optical and UV polarization of AGNs V.}

   \author{F. Marin\thanks{\email{frederic.marin@astro.unistra.fr}}}

   \institute{Universit\'e de Strasbourg, CNRS, Observatoire Astronomique de Strasbourg, UMR 7550, F-67000 Strasbourg, France}

   \date{Received ?? ??, 2018; Accepted ?? ??, 2018}

  \abstract
  {One of the main challenges for polarimetric observations of active galactic nuclei (AGN) is to
  properly estimate the amount of parasitic light that contaminates the polarization signal. 
  Removing this unpolarized flux is a complex task that has only been achieved in a couple of objects.}
  {In this fifth paper of the series, we present a new version of the Monte Carlo code STOKES  that 
  accounts for dilution by interstellar polarization and host starlight in radiative transfer modeling.}
  {We upgrade our code by including spectral energy distribution (SED) templates for a set of representative 
  host galaxies. The unpolarized light emitted by those hosts alters the observer polarization while being 
  coherently radiatively coupled to the AGN structure. We also include in our analysis tool a routine that 
  may add, depending on the user's objectives, an interstellar component.}
  {Using a generic AGN model, we illustrate how interstellar polarization and starlight dilution impact 
  the observed polarimetric signal of AGN. We apply our code to NGC~1068, an archetypal edge-on AGN and 
  demonstrate that STOKES can reproduce its SED, the expected wavelength-dependent polarimetric signatures, 
  and the observed high-angular resolution polarimetric maps. Using the flexibility of the code, we derive 
  several intrinsic parameters such as the system inclination and the torus opening angle.}
  {The new version of our publicly available code now allows observers to better prepare their observations,
  interpret their data and simulate the three-dimensional geometry and physics of AGN in order to probe 
  unresolved structures. Additionally, the radiative interaction between the host and the AGN can be used to 
  probe the co-evolution of the system.}

\keywords{Galaxies: active -- Galaxies: Seyfert -- Polarization -- Radiative transfer -- Scattering}

\maketitle


\section{Introduction}
\label{Introduction}
Spectropolarimetric observations of active galactic nuclei (AGN) have completely changed our understanding of
those objects. The discovery of broad Balmer and Fe~{\sc ii} emission lines in the polarized spectrum of NGC~1068,
with approximately the same polarization position angle and linear polarization degree that of the continuum, 
allowed \citet{Miller1983} and \citet{Antonucci1985} to reunite the spectroscopically separated type-1 and type-2 
AGN classes. A type-1 Seyfert galaxy is a lower-luminosity AGN, usually found in spiral galaxies, that often 
has two sets of emission lines in its spectrum: narrow lines such as [O~{\sc iii}]$\lambda$5007 or [S~{\sc ii}]$\lambda$6716,6731, 
with a width of several hundred of km.s$^{-1}$, and broad lines such as the Balmer series or Mg~{\sc ii}~$\lambda$2798 
lines, with widths up to 10$^4$ km.s$^{-1}$. A type-2 Seyfert galaxy lacks the broad line component, similarly to 
the so-called Narrow Line Seyfert-1s \citep{Osterbrock1989,Goodrich1989,Veron2001}. If for the latter the reason 
is unknown, it was proven thanks to spectropolarimetry that a Seyfert-1 core is hidden in probably all type-2 AGN, 
the difference being due to an orientation effect \citep{Antonucci1993}. Due to the anisotropic geometric configuration
of AGN, a type-2 observer will not see the central engine as its line-of-sight is obscured by an equatorial, optically
thick, geometrically thick dusty region. This ``torus'' is responsible for the different polarization angle we see in
type-1s and type-2s \citep{Antonucci1984}, and blocks the emission coming from the central Keplerian region of AGN 
producing the broad lines \citep{Peterson1999,Gaskell2009}. It is only thanks to spectropolarimetry that the 
dust-obscured signal of AGN cores can be revealed in type-2 Seyferts: light perpendicularly scatters along the
polar direction towards the observer and carries the broad line signatures in its Stokes parameters. 

Perpendicular scattering of optical photons leads to high polarization degrees since polarization depends on the
cosine squared of the scattering angle. We thus naturally expect highly polarized spectra for type-2 AGN as they 
are seen along the equatorial direction. Nevertheless the usual linear continuum polarization we observe is less 
than a few percent \citep[e.g.,][]{Kay1994,Ramos2016}. This is intriguing as the broad lines are detected in polarized
flux. Where are the $>>$ 10\% polarization we expect? \citet{Miller1983} showed that those high polarization degrees 
are, in fact, present but dilution by additional sources strongly reduces the observed polarization. \citet{Hiltner1949}
and \citet{Hall1949} discovered that interstellar extinction is birefringent and that the interstellar extinction 
increases greatly towards the ultraviolet (UV). Interstellar polarization is to be accounted for, since the trajectory
of photons through the local galactic medium is impacted by the alignment of dust grains and produces a foreground
contamination. Hence, any observations of distant sources might be altered by interstellar polarization. This can 
be estimated by looking at the polarization of Galactic stars in the same field of view than the AGN \citep{Miller1990}, 
or by using plots of the $E$-vector of polarization over the entire sky \citep{Mathewson1970}. On the other hand, 
starlight from the host galaxy is a major contributor to the contamination of polarized spectra of type-2 AGN: since
the nucleus is hidden, a large fraction of the observed flux comes from stellar sources and the final polarization 
decreases. This is less problematic for the case of type-1 AGN as the central engine is directly visible through 
the torus funnel and radiation from the accretion disk dominates the host starlight. However, for Seyfert-2s, it becomes
a major challenge to estimate the true scattered component of light. It is possible to remove the starlight 
contribution using a galaxy template superimposed on high-resolution polarized spectra. By carrying out 
subtractions at various flux ratios between the template host and the polarized spectrum, it is possible 
to find a flux ratio that provides a simultaneous cancellation of all the stellar features. This was 
the technique used by \citet{Miller1983}, who found that the true continuum polarization of NGC~1068 is
polarized at $P$ = 16\% $\pm$ 2\%. The wavelength-independence of the polarization continuum was also a proof 
for the dominance of Thomson scattering in the optical band. However, not all archival polarimetric observations 
were achieved in high resolution spectropolarimetric modes. To correct  for interstellar and host dilution 
the observed narrow band and filter-integrated polarimetric observations, reverse engineering is needed. 

This is the purpose of our upgrade. We aim at providing a simulation tool that can fit any near-infrared, 
optical and ultraviolet polarimetric observations of AGN and provide a good estimation of its true 
scattered polarization. To do so, we include a variety of galaxy templates to be accounted for in AGN 
simulations. The flux ratio between the AGN central engine and the host can be arbitrarily chosen by 
the user to fit the data and retrieve the true polarization. We present in Sect.~\ref{STOKES} an 
overview of the current code version and explain the upgrade we included in the newest version. We 
show in Sect.~\ref{Modeling} several examples of polarimetric simulations, both in photometry, spectropolarimetry 
and polarization imaging, to highlight the necessity of accounting for interstellar polarization and 
starlight contribution. We discuss the potential future use of our code in Sect.~\ref{Discussion} and 
conclude our paper in Sect.~\ref{Conclusions}.

\section{STOKES upgrade}
\label{STOKES}
{\sc stokes} is a Monte Carlo code that was first presented in \citet{Goosmann2007}. Its main goal is to 
reproduce the observed near-infrared, optical and ultraviolet polarization of AGN. The most important part 
of this radiative transfer code is that it accounts for radiative coupling between the different three-dimensional
emitting and scattering regions. Multiple scattering has a profound impact on the net polarization we observe
and several specific features seen in polarimetry cannot be reproduced without multiple scattering (see, e.g., 
\citealt{Young2000,Smith2002} who numerically reproduced the polarization variations observed in broad emission
lines). Multiple scattering is thus essential, and it is also the backbone of our new numerical improvements.

\subsection{Overview of the code performances}
\label{STOKES:code}
The previous version (v1.2) of {\sc stokes} is able to handle photon emission with a variety of 
spectral shapes (power-law, thermal emission, ...) originating from any number of geometrical sources that 
can be located anywhere in the three-dimensional environment. The photons then travel in straight lines (relativistic 
effects are yet to be included in the vicinity of the central black hole) until they encounter a medium whose size,
composition, temperature and velocity can be fully parametrized by the user. Depending on the photon energy and 
the opacity of the matter, radiation can travel through the medium, be scattered, absorbed and potentially re-emitted
several times. The code registers the photon time delay due to multiple scattering and follows each photon until
it is absorbed or escapes from the model region. A web of virtual detectors (scaled in cosine along the polar direction) 
registers the photons at all viewing angles (including potential azimuthal bins). The photon polarization is stored
using the Stokes parameters I, Q, U and V. The I parameter stands for the intensity, Q and U are used to characterize
the linear polarization, and V characterizes the circular polarization. 

The code's output is given in a specific format that can be processed with the {\sc analyze} tool that is provided 
with the code (http://www.stokes-program.info/ ). The {\sc analyze} routine computes the total flux, the total, linear 
and circular polarization degrees ($P$ = $P_{\rm lin}$ + $P_{\rm circ}$), the polarization position angle $\Psi$ 
and the polarized flux (total flux times $P$). A convention used since the first paper of our series is to 
consider the polarization to be parallel when the $\vec E$-vector is aligned with the projected symmetry axis of the model 
(i.e. $\Psi$ = 90$^\circ$). We sometimes denote the difference between parallel and perpendicular polarization by 
the sign of the polarization percentage. $P$ is negative if $\Psi = 90^\circ$ and $P$ is positive if $\Psi=0^\circ$. 

During the past decade, the code has been improved in many ways. It is now equipped with an imaging routine that can 
have any required spatial resolution \citep{Marin2012}, it can also account for fragmented regions made of several 
thousands spherical clumps \citep{Marin2015}, and it coherently computes the time-delays and cross-correlation 
of light between the total and unpolarized fluxes \citep{Rojas2018}. The code has been implemented with faster 
computing routines to allow swifter runs. An overview of the science cases were {\sc stokes} was applied 
is presented in \citet{Marin2014}.

\subsection{Interstellar polarization}
\label{STOKES:ISP}

\begin{figure}
    \includegraphics[trim = 0mm 0mm 0mm 0mm, clip, width=8.9cm]{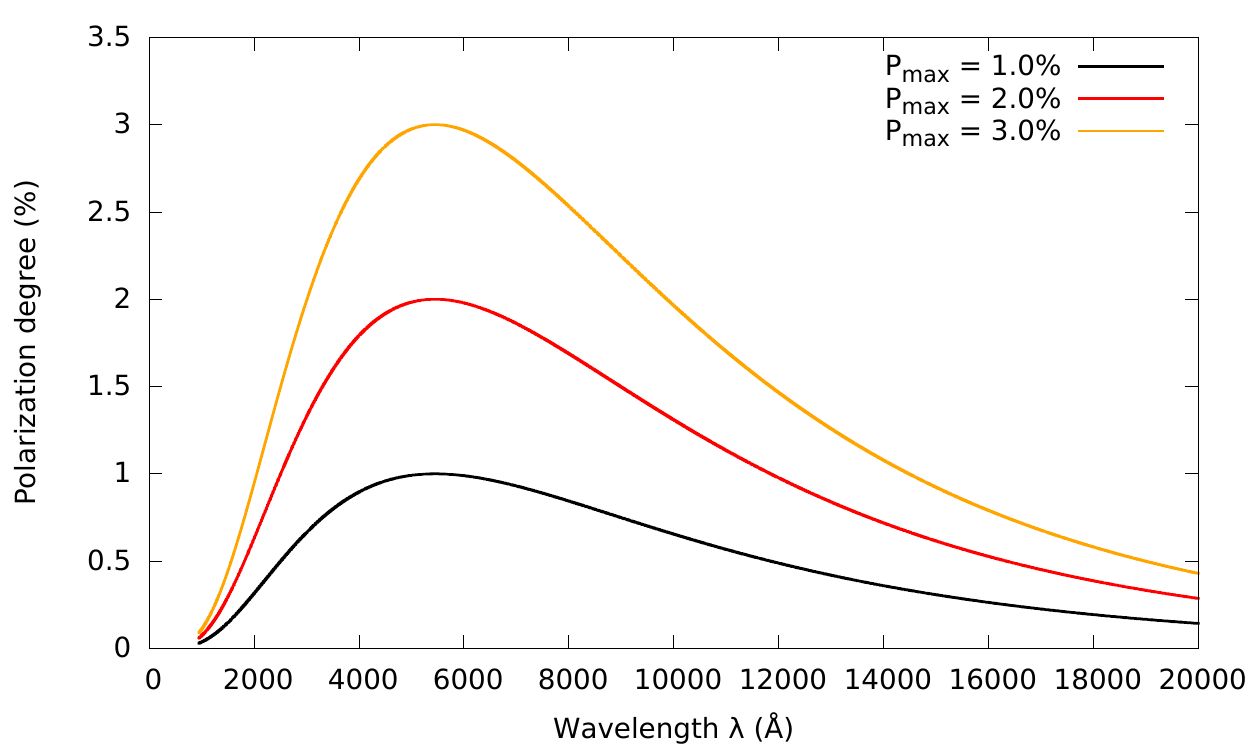}
    \caption{Examples of ultraviolet-to-infrared interstellar
	      polarization levels for different values of
	      $P_{\rm max}$. The value of $\lambda_{\rm max}$ 
	      is fixed to 5450\AA.}
    \label{Fig:ISP}
\end{figure}

\citet{Hiltner1949} and \citet{Hall1949} have shown that light reaching us through the interstellar medium is polarized.
In the interstellar medium, the dust grains, which are known to be elongated \citep{Hildebrand1995}, can be aligned by 
the interstellar magnetic field. The aligned grains block the light waves that are polarized parallel to the
long axis of the grains, and let the polarized waves pass perpendicularly to the long axis of the grains. This dichroic 
extinction is responsible for an interstellar polarization that can reach up to 5\% \citep{Mathewson1970}. This polarized 
signal can easily dilute or overwhelm the polarization signal of type-1 AGN, which is known to be weak (often less than 2\%) 
due to their symmetric morphology \citep{Marin2012}. It is particularly important to include this feature in 
{\sc stokes} to allow the user to check whether the polarimetric signal observed is truly due to the AGN. As we will
show in Sect.~\ref{Modeling}, constraints on the inclination angle of the AGN can also be deduced thanks to comparisons
between observations and modeling that accounts for interstellar polarization.

To include interstellar polarization, we follow the seminal work of \citet{Serkowski1975}. The authors investigated 
the wavelength dependence of interstellar linear polarization for $\sim$180 stars with a variety of photometric filters.
By doing so, they produced a well-known empirical formula:

\begin{equation}
   P(\lambda) = P_{\rm max} \exp\left(-1.15 \ln^2(\frac{\lambda_{\rm max}}{\lambda})\right),
\end{equation}

where $\lambda_{\rm max}$ is the wavelength at which the maximum interstellar linear polarization $P_{\rm max}$ occurs.
From observations, \citet{Serkowski1975} determined that $\lambda_{\rm max}$ = 5450\AA. We illustrate different interstellar 
polarization signatures in Fig.~\ref{Fig:ISP} for different values of $P_{\rm max}$. The optical band suffers the most 
from this parasitic component and observations in the ultraviolet or infrared bands must be favored. The empirical 
law of \citet{Serkowski1975} is now implemented in {\sc analyze}. The user has to enter the values of $\lambda_{\rm max}$,
$P_{\rm max}$ and $\Psi_{\rm ISP}$. The parameter $\Psi_{\rm ISP}$ was not investigated by \citet{Serkowski1975} but it has some 
importance since it can alter the observed polarization position angle of the AGN polarized flux, even when the magnitude 
of interstellar polarization is small. \citet{Hoffman2005} studied how such uncertainties may affect the observed 
polarization angle and found that the intrinsic position angle differs from the observed one by

\begin{equation}
   \Delta\theta_{\rm max} = 0.5\sin^{-1}\left(\frac{P_{\rm ISP}}{P_{\rm obs}}\right).
\end{equation}

We thus decided to also account for the variation of the total polarization position angle by introducing in our code
the parameter $\Psi_{\rm ISP}$ that can be easily derived from the previous equation.

\subsection{Host starlight}
\label{STOKES:host}

\begin{figure}
    \includegraphics[trim = 5mm 5mm 5mm 30mm, clip, width=10.5cm]{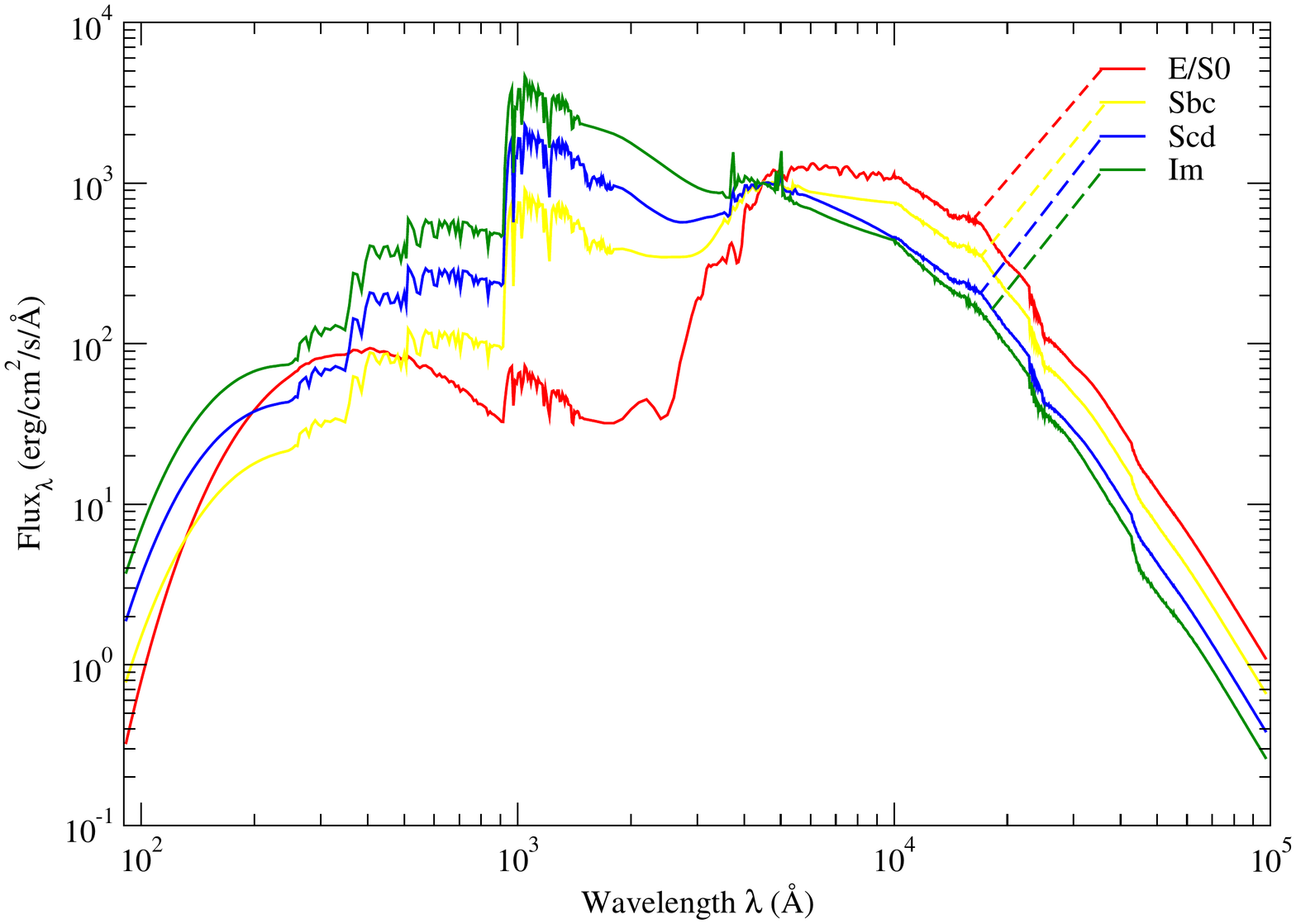}
    \caption{Templates of four classes of host galaxy
	    extracted from \citet{Bruzual1993} and 
	    extended towards the ultraviolet and infrared 
	    bands by \citet{Bolzonella2000}. The four 
	    templates are: an elliptical/lenticular 
	    galaxy (E/S0, in red), a spiral galaxy with 
	    very loose spiral arms (Sbc, in yellow), 
	    a spiral galaxy (Scd, in blue), and an
	    highly irregular galaxy (Im, in green).}
    \label{Fig:Host}
\end{figure}

Dilution by starlight is the major contributor to the weakening of the observed AGN polarization. Galaxies
contain large amounts of hot gas that are detected thanks to intense emission lines, and stars are responsible
for the continuum spectrum and the absorption lines. There is a vast collection of galaxy classes that are 
divided into groups based on their visual appearance \citep{Hubble1936,Vaucouleurs1959,Sandage1975,Bergh1998}.
However AGN are mostly found in the heart of early-type spiral galaxies (Sa-Sb, \citealt{Heckman1978}). Large
surveys of well-resolved nearby AGN confirmed this trend but also demonstrated that Seyfert host galaxies
often show signs of morphological irregularities, most probably caused by tidal interactions \citep{Simkin1980,MacKenty1990}.
Not all Seyferts show a large-scale stellar bar within the spiral host galaxy \citep{Deo2006} and \citet{Kelm2001}
have shown that about 70\% of both Seyfert-1 and Seyfert-2 classes are preferably detected in spirals.
\citet{Chen2017} also discovered that Seyfert-1 galaxies are mainly located in bulge-dominant galaxies while 
type-2s are more often detected in disk galaxies. However, since both classes are dominated by large bulge 
fractions, late type host galaxies cannot be excluded. For this reason, we included in {\sc stokes} the possibility
to use a variety of host galaxy templates that cover a large fraction of the galaxy morphological classification.
We extracted the templates of four galaxy types (E/S0, Sbc, Scd, and Im) from the public code {\sc hyperz} 
\citep{Bolzonella2000} and included them in our code. The host templates, see Fig.~\ref{Fig:Host}, cover the 
90 -- 100,000~\AA~band. They are soft-coded in {\sc stokes}, which means that the user can erase the source file 
and replace the input spectrum with its own galaxy spectral energy distribution (SED). An automatic procedure rebins 
the host spectrum according to the required resolution set by the user. Since stars principally emit unpolarized 
or mildly polarized light, and since the vectorial superposition of the many polarization angles from all the stars 
of the galaxy will be null, the common assumption of unpolarized host radiation is used in our code. However,
scattering of starlight by dust, molecules and electrons in the galactic medium may produce an observable 
amount of polarization that depends on the orientation of the host plane \citep{Simmons2000,Stil2009}. In our model, it 
is therefore possible to fill the host galaxy region with dust and/or electrons to re-create such conditions. 
Finally, the fraction of host starlight with respect to the AGN intrinsic flux, integrated over the considered 
waveband, is required before starting the code.

\subsection{Quantitative assessment}
\label{STOKES:test}

\begin{figure*}
    \centering
    \includegraphics[trim = 0mm 0mm 0mm 0mm, clip, width=20cm]{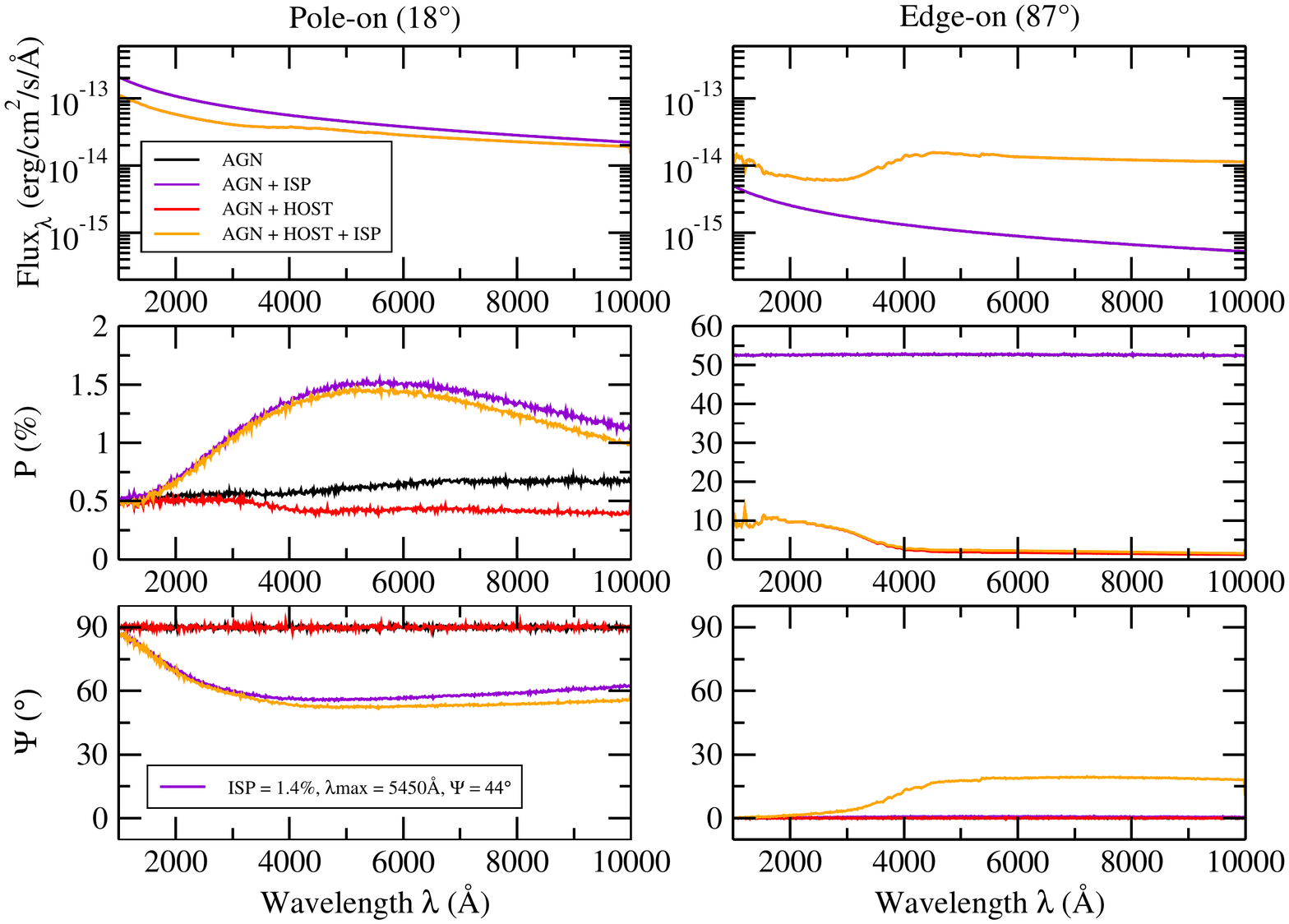}
    \caption{Impact of interstellar polarization (ISP) and starlight
	    dilution on the AGN total flux (top), polarization
	    degree (middle) and polarization angle (bottom). Two 
	    different inclinations are shown: a polar orientation
	    (left) and an equatorial inclination (right). Four AGN
	    models are shown. In black is an AGN model without 
	    parasitic light (the kind of model that {\sc stokes}
	    was able to produce until this paper). In violet is 
	    a model that accounts for ISP, in red a model that accounts
	    for host starlight only, and in orange an AGN model
	    that accounts for both the host and ISP. The interstellar
	    polarization is parametrized with the following numerical values: 
	    $P_{\rm max}$ = 1.4\%, $\lambda_{\rm max}$ = 5450~\AA,
	    and $\Psi_{\rm ISP}$ = 44$^\circ$.}
    \label{Fig:test}
\end{figure*}

We illustrate how interstellar polarization (ISP) and starlight dilution can alter the observed total flux
and polarization of a cosmic source in Fig.~\ref{Fig:test}. We simulated four different AGN observations
that account for 1) only the AGN flux, 2) the AGN flux plus ISP, 3) the AGN flux plus the host starlight, 
and 4) the AGN flux plus the host and the ISP. We set the ratio of AGN photons to host photons to 50\% for 
this experiment. The top row of Fig.~\ref{Fig:test} shows that the ISP has no impact on the observed
AGN flux (the black and violet lines are superimposed, similarly to the red and orange lines). The host 
starlight has very little influence on the observed spectrum of type-1 objects, since emission from the 
accretion disk dominates. On the other hand, we see that the host overshadows most of the AGN flux for 
type-2 objects. The ultraviolet band is less affected due to the dimming of starlight photons in this 
waveband but the optical and near-infrared bands are completely dominated by the host. Looking at the 
polarization degree (Fig.~\ref{Fig:test}, middle panels), we see that type-1 polarization is only
marginally affected by starlight dilution in the optical band. The ISP, however, can completely cover 
the AGN polarization. At equatorial inclinations the presence of ISP is completely overshadowed by the 
high polarization degrees due to perpendicular scattering of disk photons in the polar outflows. Without
host contamination $P$ can be as high as 53\%. When starlight is added, the polarization degree is strongly 
weakened. It is no longer constant and shows the highest values in the ultraviolet band, such as observed 
by \citet{Code1993}. The overall polarization degree is strongly affected and reduced to $P$ less than
10\% in the ultraviolet, and less than 3\% in the optical. Those values correspond to the numerous polarimetric 
observations of type-2s achieved by \citet{Kay1994}. Finally, the polarization position angle is not 
affected by the host starlight at polar inclinations, but the ISP can induce a wavelength-dependent variation 
(Fig.~\ref{Fig:test}, bottom panels). For equatorial inclinations, only the combination of starlight
dilution plus interstellar polarization may affect $\Psi$.

Overall, we have seen that removing both the ISP and the host contamination is necessary to uncover
the true polarization originating from the AGN. While correcting for interstellar polarization 
is achievable (yet it requires observations of standard stars), accounting for host starlight 
is more tricky. Radiative transfer models are thus needed to uncover the true scattered polarization 
thanks to retro-engineering.

\section{Examples of NGC~1068 modeling}
\label{Modeling}
In order to demonstrate how our newest code version can help to disentangle the true AGN polarized 
light from parasitic light, we run {\sc stokes} for a realistic test case. We focus on one specific 
target for this study: NGC~1068. This archetypal Seyfert-2 galaxy has been the most observed in 
terms of polarization and archival data allow us to efficiently compare our simulations to observations. 
Details of the model are listed in Tab.~\ref{Tab:Model}. Each free parameter is described and
referenced in the consecutive subsections. In addition, since it is an AGN seen at a relatively large 
inclination (70 -- 80$^\circ$, \citealt{Honig2007,Fischer2013,Fischer2014,Marin2016}), it is an ideal 
target to evaluate the fraction of polarization dilution in dust-obscured AGN.

\begin{table}
  \centering
  \begin{tabular}{lr}
    \textbf{Continuum source} & ~ \\  
    \hline
    Punctual isotropic power-law SED: & $F_{\rm *}~\propto~\nu^{-1}$ \\
    ~ & ~ \\
    \textbf{Accretion flow} & ~ \\  
    \hline    
    Inner and outer radii: & 0.001 -- 0.1~pc \\
    Half-opening angle (from the eq. plane): & 20$^\circ$ \\
    Composition and optical depth: & electrons ($\tau_{\rm V}$ = 1) \\
    ~ & ~ \\
    \textbf{Circumnuclear region} & ~ \\  
    \hline    
    Inner and outer radii: & 0.1 -- 5~pc \\
    Half-opening angle (from the eq. plane): & 45$^\circ$ \\
    Composition and optical depth: & dust ($\tau_{\rm V} \ge$ 50) \\    
    ~ & ~ \\
    \textbf{Biconical polar outflows} & ~ \\  
    \hline    
    Inner and outer radii: & 0.1 -- 100~pc \\
    Half-opening angle (from the eq. plane): & 45$^\circ$ \\
    Composition and optical depth: & electrons ($\tau_{\rm V}$ = 1) \\      
    ~ & ~ \\
    \textbf{Host galaxy} & ~ \\  
    \hline    
    Template: & Sbc \\
    Inner and outer radii: & 5 -- 3,000~pc (truncated) \\
    ~ & ~ \\
    \textbf{Interstellar polarization} & ~ \\  
    \hline    
    Polarization degree: & 0.5\% \\
    Polarization angle: & 0$^\circ$ \\    
  \end{tabular}
  \caption{Components of the NGC~1068 model, its host
	   galaxy and the interstellar polarization. 
	   See text in Sect.~\ref{Modeling:sed}, 
	   \ref{Modeling:continuum} and \ref{Modeling:maps}
	   for details and references.}
  \label{Tab:Model}
\end{table}

\subsection{Spectral energy distribution}
\label{Modeling:sed}

\begin{figure}
    \includegraphics[trim = 5mm 5mm 0mm 10mm, clip, width=10.1cm]{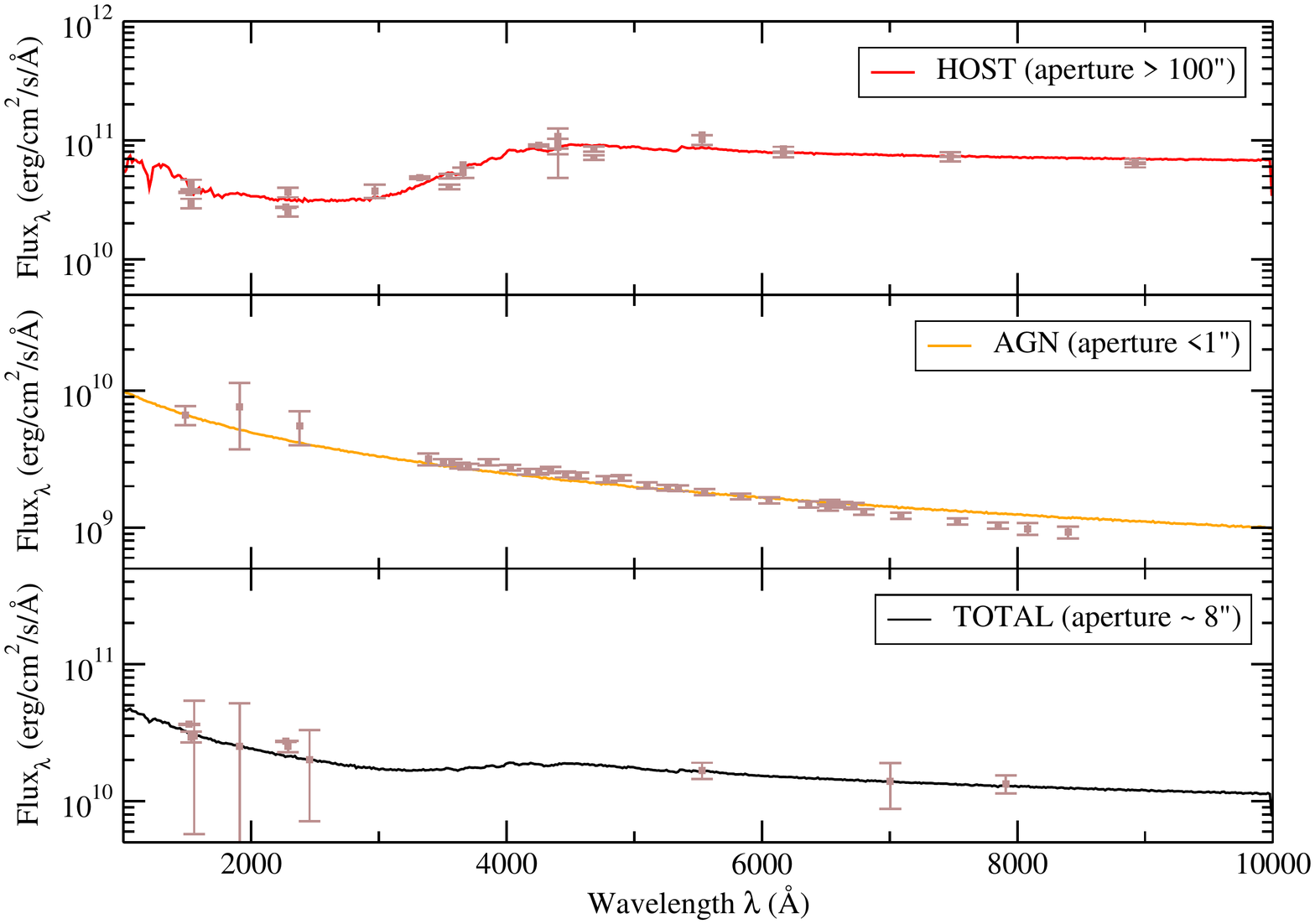}
    \caption{Decomposition of the ultraviolet, optical and near-infrared 
	    SED of the Seyfert-2 galaxy NGC~1068, including multi-aperture
	    flux measurements taken from the NASA/IPAC Extragalactic Database
	    (NED). Dust re-emission by the AGN circumnuclear region is not 
	    accounted for as it only becomes preponderant at $\lambda >$ 1$\mu$m.}
    \label{Fig:SED}
\end{figure}

We start by modeling the 1000 -- 10,000~\AA~SED of NGC~1068 (Fig.~\ref{Fig:SED}). To represent the host galaxy 
contribution, we use the Sbc template presented in Sect.~\ref{STOKES:host}. This template is consistent with the 
morphological classification of the host of NGC~1068, such as discovered by \citet{Balick1985}. The authors also 
highlighted the importance of starburst activity in the SED of NGC~1068 that impacts the observed fluxes at wavelengths
larger than 1$\mu$m. The resulting host SED nicely fit the fluxes observed with large aperture detectors that are 
listed in the NASA/IPAC Extragalactic Database, see the first panel of Fig.~\ref{Fig:SED}. The AGN continuum is 
modeled using an isotropic power-law spectral energy distribution $F_{\rm *}~\propto~\nu^{-\alpha}$ with $\alpha = 1$.
This continuum is the scattered component of the intrinsic flux that is blocked along the equatorial plane by an 
obscuring dust screen, usually refereed to as a ``torus''. The scattered continuum originates from polar scattering 
of disk photons. This simple model allows us to reproduce the observed AGN flux that can only be detected with small 
aperture instruments ($\le$ 1$"$), see the second panel of Fig.~\ref{Fig:SED}. Finally, the combination of the two 
components allow us to reproduce the observed SED of NGC~1068 at apertures close to 8$"$. Assuming a distance of 
$\sim$ 14.4~Mpc, 1$"$ corresponds to 72~pc; thus an 8$"$ aperture isolates the first 576~pc around the nucleus, 
where both the host and the AGN contribute to the observed flux. Hence, because of the inclusion of host starlight 
templates, {\sc stokes} is now possible to properly compute and reproduce the observed SED of AGN, while accounting 
for the aperture effects.

\subsection{Continuum linear polarization}
\label{Modeling:continuum}

\begin{figure}
    \includegraphics[trim = 5mm 5mm 0mm 10mm, clip, width=10.1cm]{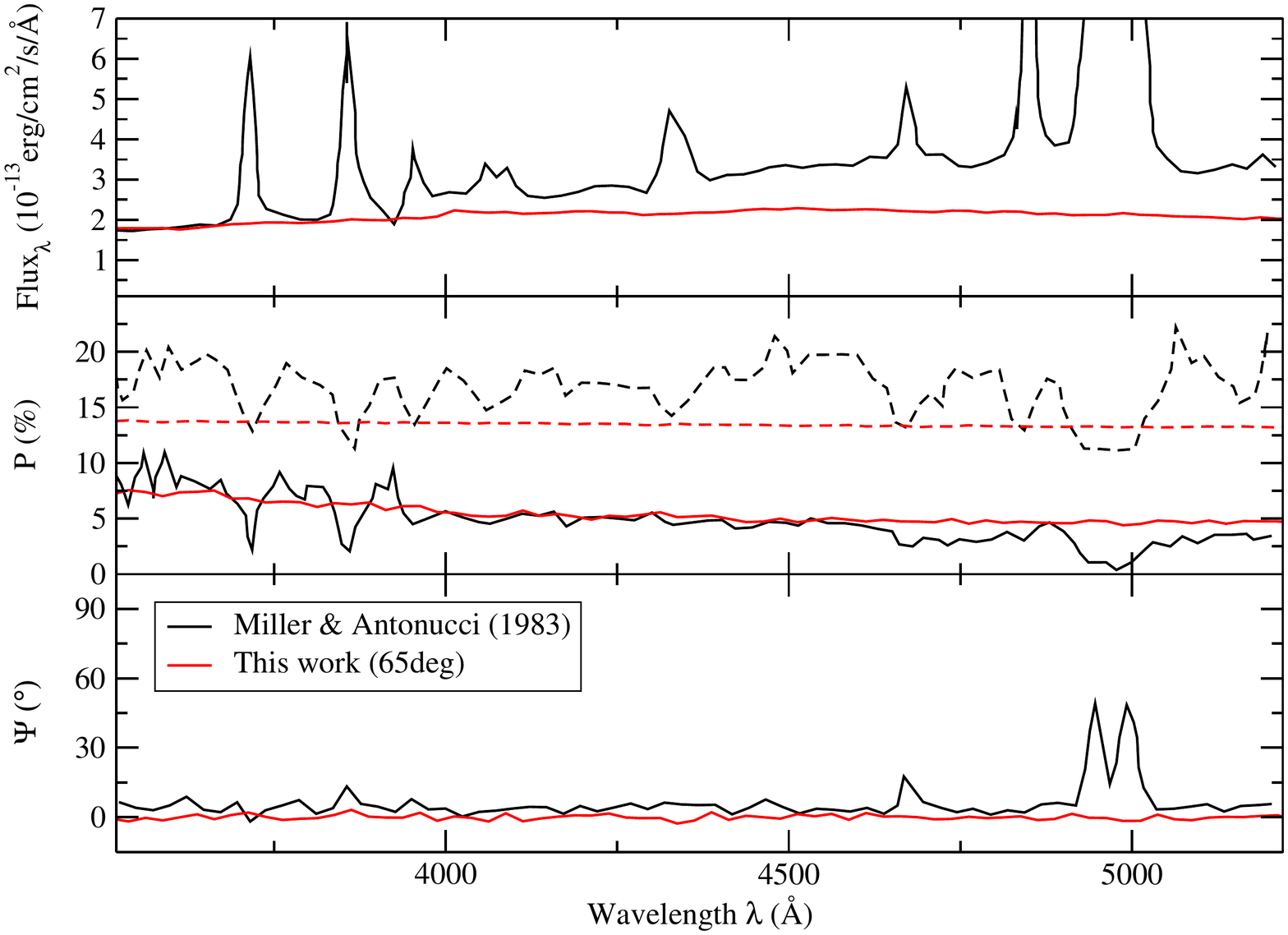}
    \caption{Total flux, linear polarization degree $P$ and 
	     polarization position angle $\Psi$ resulting from 
	     the modeling of NGC~1068 including ISP and starlight 
	     dilution by a Sbc host galaxy. In black are
	     the observations by \citet{Miller1983} and 
	     in red are the Monte Carlo simulations.
	     The solid line corresponds to the model 
	     with starlight dilution and the dashed line 
	     is the polarization degree without host 
	     contamination.}
    \label{Fig:spectroNGC1068}
\end{figure}

We pursue our study of NGC~1068 by investigating its continuum polarization. To do so, we construct a physical 
model of the AGN by setting a power-law continuum source at its center with the same characteristics as the ones described in 
Sect.~\ref{Modeling:sed}. Around this central source, we implement a scattering disk that is representative 
of the accretion flow between the torus and the accretion disk \citep{Smith2002}. This structure is marginally
optically thin ($\tau_{\rm V}$ = 1), fully ionized and extends from 0.001 to 0.1~pc \citep{Marin2012}. It has a
flared structure with a half-opening angle of 20$^\circ$ from the equatorial plane (\textit{ibid.}). At larger 
distances from the source, we set up the dusty circumnuclear region using a flared disk filled with Milky Way 
dust. The structure is optically thick ($\tau_{\rm V} \ge$ 50) and extends from 0.1 to 5~pc. The inner 
radius is set by the dust sublimation radius \citep{Barvainis1987,Kishimoto2007} and the outer rim corresponds
to the expected size of AGN tori \citep{Siebenmorgen2015,Marin2015}. The half-opening angle of the torus is fixed
to 45$^\circ$ \citep{Sazonov2015}. Finally, along the polar direction, biconical outflows are collimated by 
the torus funnel. They extend up to 100~pc and they are filled with electrons and their radial optical depth is 
fixed to $\tau_{\rm V}$ = 1 \citep{Osterbrock1989}. See Tab.~\ref{Tab:Model} for a summary of the model 
parameters. It is a generic model based on the Unification Scheme presented by \citet{Antonucci1993}, yet such 
modeling has been very successful in reproducing observational polarimetric measurements \citep{Marin2012,Marin2015}. 
Our goal is not to perfectly reproduce the spectropolarimetric signal of NGC~1068, but rather show whether the 
Unified Scheme gives a consistent comparison between observations and simulations, and to show how {\sc stokes} 
can help to fit data. Around the AGN, we added a host galaxy using the Sbc template presented in the previous 
section. The flux ratio between the two radiation sources is fixed according to the estimation made by \citet{Miller1983}
who concluded that ``no more than 15\% of the light at 4000\AA~could come from stars earlier than type F''. Radiation
from the host may scatter onto the AGN, coupling the radiation field in a consistent way. Finally, an interstellar 
polarization of 0.5\% at 0$^\circ$ was added on the detector plane, following the measurements of \citet{Miller1983} 
and the all-sky polarization maps of \citet{Mathewson1970}.

We show in Fig.~\ref{Fig:spectroNGC1068} the resulting continuum flux, polarization degree and polarization angle 
produced by {\sc stokes} for the same model but using the old and new versions of the code. The observed 
spectra (in black) are extracted from Miller's publication using WebPlotDigitizer, a polyvalent and free software 
to extract spectra from old astronomical publications \citep{Marin2017}. Since {\sc stokes} records the net 
polarization at all inclinations, we tested all the spectra. The best fit between observations and our model
is obtained for an inclination of 67.5 $\pm$ 2.5$^\circ$, shown in red in Fig.~\ref{Fig:spectroNGC1068}. We see that 
the continuum level is found to be very similar to the observed continuum flux, albeit with small differences at longer 
wavelengths. This is mainly due to the absence of emission lines whose wings contribute to the continuum. There is 
a great agreement between the observed starlight-contaminated polarization degree $P$ and the continuum polarization 
simulated with {\sc stokes}, both in percentage and wavelength-dependence. This agreement could not be obtained
with the previous version of the code since dilution was not accounted for. The observed polarization level was too high 
in comparison to observations. Our new version of {\sc stokes} allows us to subtract the starlight component to the 
linear polarization, enabling us to reproduce the scattering-induced polarization estimated by \citet{Miller1983}.
The wavelength-independence of $P$ is due to electron scattering and nicely reproduces the observational work of the 
authors. The polarization position angle (compared to the position angle of the radio structure in NGC~1068) is 
perpendicular (hence equal to 0$^\circ$ using our polarimetric convention), similarly to the observed value.

Our modeling has demonstrated that it is possible to reproduce both the total and polarized components of AGN
using the latest code implementations. Such work was not previously possible without using crude first-order
approximations for flux and polarization dilution. Now, not only can the code fit data with a physical model, but it 
is also possible to derive the inclination angle of the AGN. Using a simple yet reliable AGN model, we derived an inclination 
angle for the AGN core that is consistent with previous estimations \citep{Marin2016}. However we found a small 
degeneracy between the inclination angle and the half-opening angle of the dusty circumnuclear structure. Varying 
the height of the dust component (with respect to the equatorial plane) impacts the inclination angle at which the 
total flux and polarization degree fit the observational data the best. The degeneracy is not linear and is only 
revealed due to multiple scattering. For each AGN, a set of models must be run in order to evaluate the uncertainties 
on the inclination angle and torus geometry. This is beyond the scope of our current paper that aims to present
the numerical tool needed for such an extensive analysis.

\subsection{Polarimetric maps}
\label{Modeling:maps}

NGC~1068 was also extensively observed in imaging polarization thanks to the Hubble Space Telescope (HST) 
and various other telescopes. Recently \citet{Gratadour2015} undertook high angular resolution polarization 
imaging of the first 5$"$ around NGC~1068. Using the infrared camera IRDIS installed in the SPHERE instrument
on the Very Large Telescope, \citet{Gratadour2015} achieved 60~mas H and K' band polarimetric maps of 
the central region of NGC~1068. Similar, yet less resolved, imaging polarimetric campaigns were achieved 
by, e.g., \citet{Capetti1995a,Capetti1995b} and \citet{Weinberger1999}. In order to demonstrate the 
code capabilities to reproduce such observations, we run {\sc stokes} with a 200x200 spatial resolution binning
covering the first 300~pc around the nucleus. We implemented a de~Vaucouleurs profile for the emission of 
the central parts of the host galaxy since the bugle region of spiral galaxies can be well approximated by 
an elliptical model \citep{Vaucouleurs1948}. The surface brightness of the host is then modeled using the 
following equation:

\begin{equation}
   I(R) = I(R_{\rm e}) \exp\left(-b (\frac{R}{R_{\rm e}})^{1/n}-1\right),
\end{equation}

with $R_{\rm e}$ the radius of the isophote containing half the galaxy luminosity, $b$ = 1.999$n$-0.327 and $n$ = 4. 
Following the NGC~1068 host galaxy observations of \citet{Tanaka2017} we fix $R_{\rm e}$ = 1,500~pc. The host 
galaxy size ranges from 5 to 3,000~pc in our model and the fraction of starlight in the observed total continuum
flux is set to 0.82 at 5075~\AA, such as measured by \citet{Miller1983}. Finally, we fill the host disk with 
dusty grains so that the vertical density structure is optically thin.

\begin{figure*}[ht] 
   \begin{minipage}[b]{0.5\linewidth}
      \centering
      \includegraphics[trim = 0mm 0mm 0mm 5mm, clip, width=9cm]{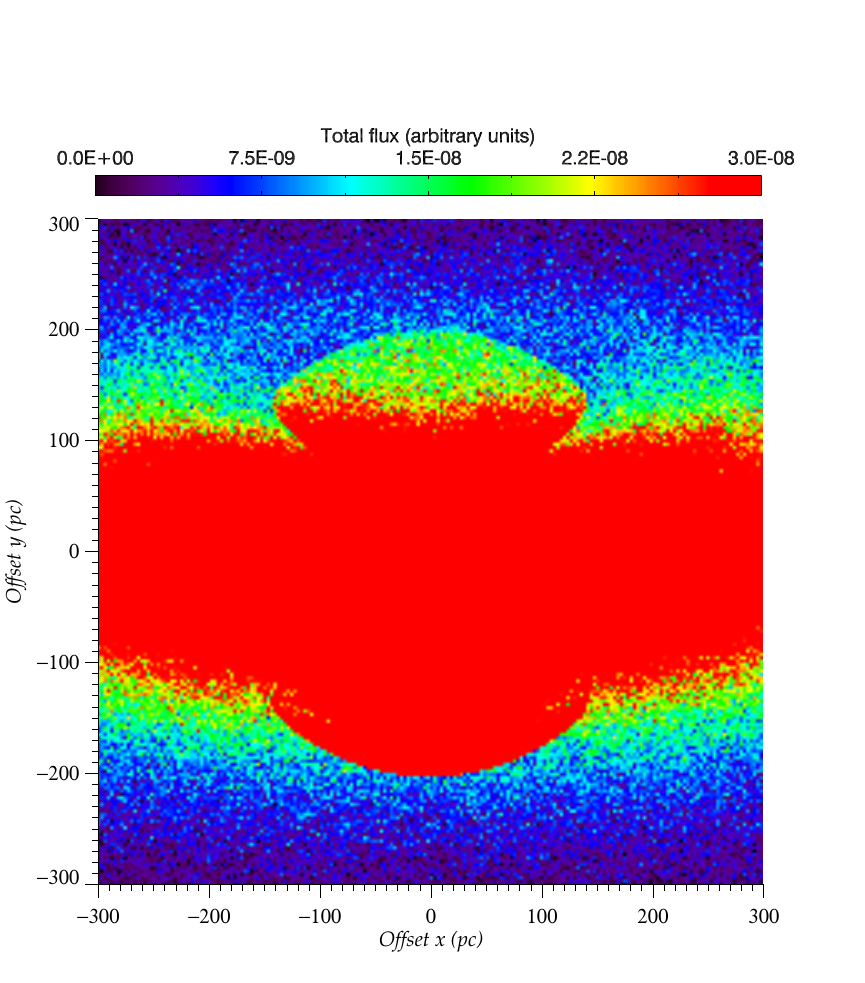} 
   \end{minipage} 
   \begin{minipage}[b]{0.5\linewidth}
      \centering   
      \includegraphics[trim = 0mm 0mm 0mm 5mm, clip, width=9cm]{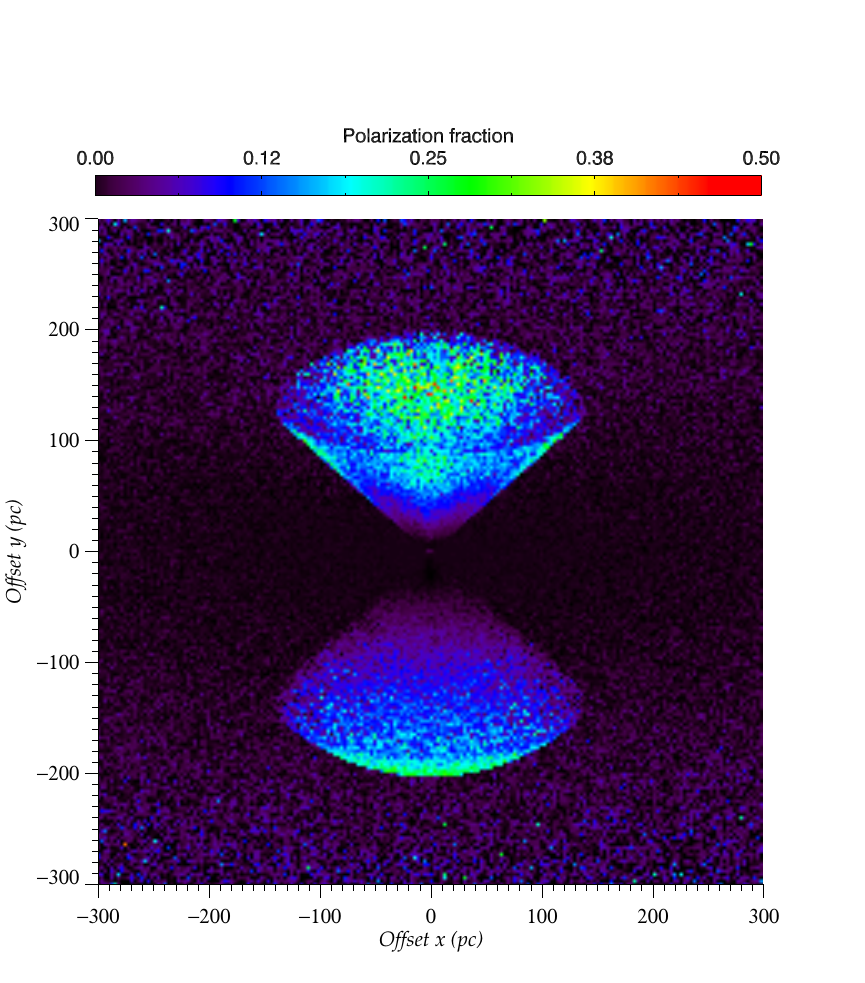} 
   \end{minipage}
   \begin{minipage}[b]{0.5\linewidth}
      \centering
      \includegraphics[trim = 0mm 0mm 0mm 10mm, clip, width=9cm]{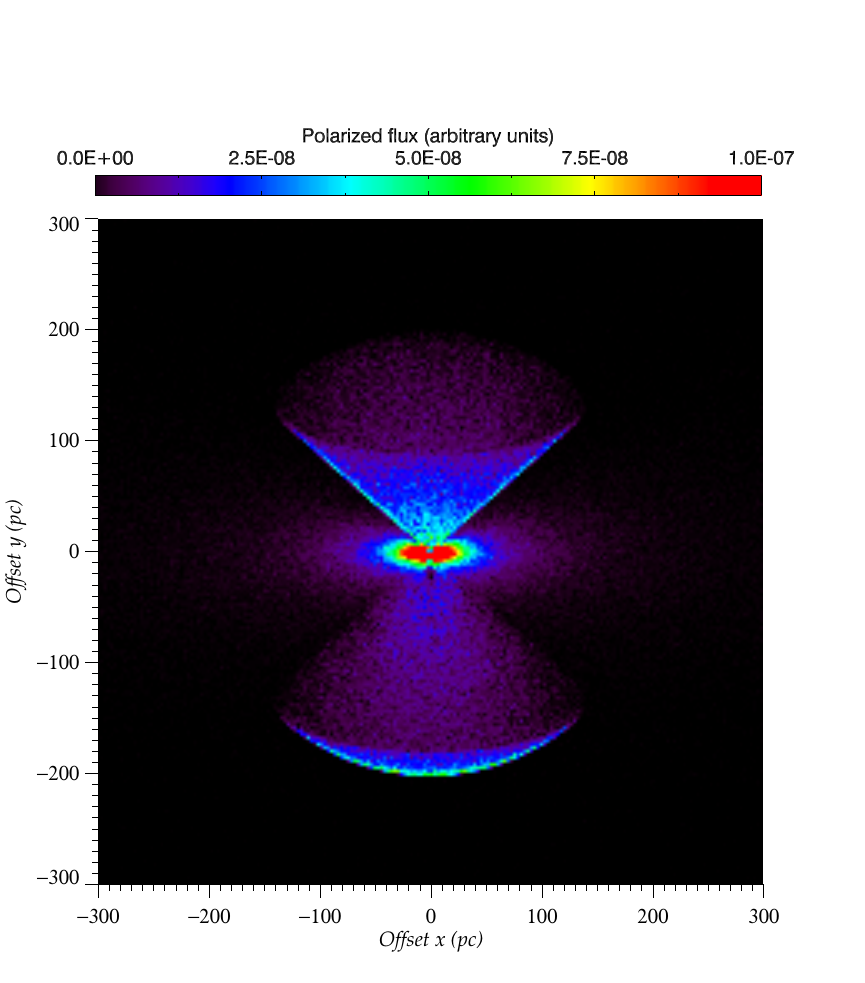} 
   \end{minipage} 
   \begin{minipage}[b]{0.5\linewidth}
      \centering   
      \includegraphics[trim = 0mm 0mm 0mm 10mm, clip, width=9cm]{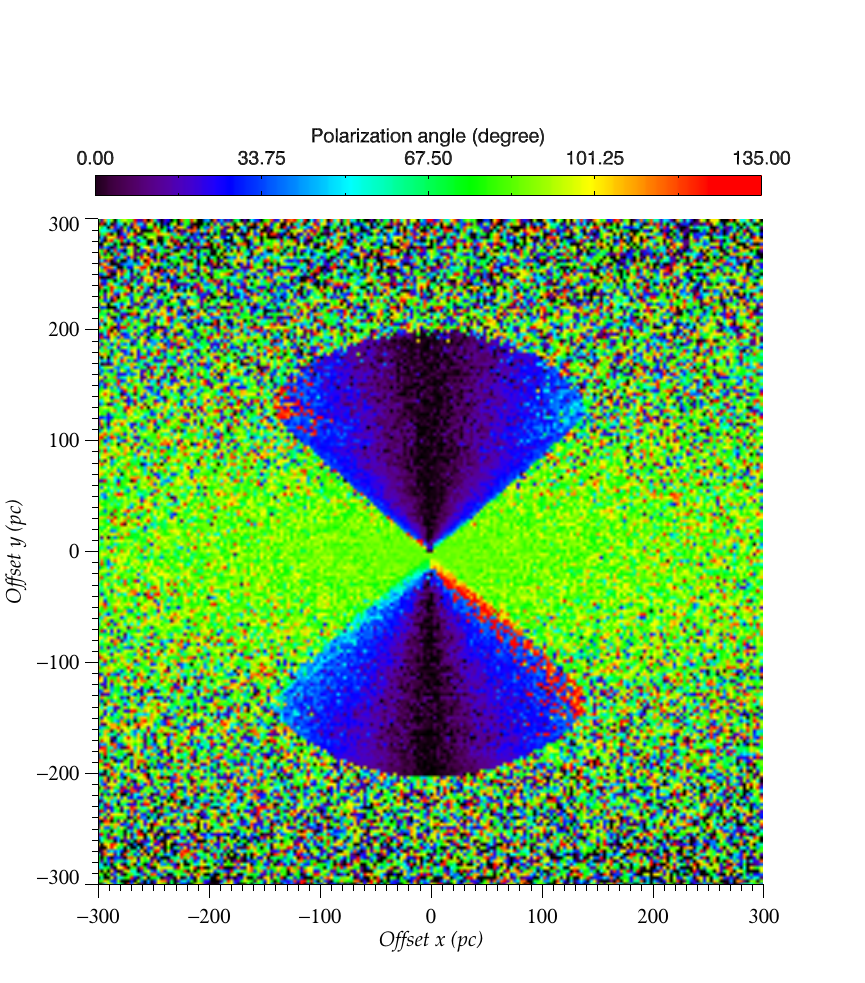} 
   \end{minipage}    
  \caption{Monochromatic ($\lambda$ = 5000~\AA) polarization 
	  maps of NGC~1068. Top-left: total flux, top-right:
	  fraction of polarization, bottom-left: polarized 
	  flux, bottom-right: polarization angle. In the case 
	  of the total and polarized fluxes the color bar is 
	  in arbitrary units.}
  \label{Fig:Maps}
\end{figure*}

We present the monochromatic ($\lambda$ = 5000~\AA) polarization maps of NGC~1068 in Fig.~\ref{Fig:Maps}. 
The top-left map displays the spatial distribution of the total flux. The core of the picture is saturated 
due to the strong central emission of the host galaxy. The de~Vaucouleurs emission profile is clearly 
visible, with decreasing fluxes observed at larger distances from the central bins. The top and bottom parts 
of the polar outflows are almost detected as host photons from the background are scattered away from the observer's
line-of-sight by electrons in the polar wind, hence contrasting the sharp borders of the outflows. The rest
of the AGN cannot be detected as the image is entirely dominated by the central photon flux. Compared 
to real total flux images of the central parts of NGC~1068 (see Fig.~1 in \citealt{Gratadour2015}) our 
simulations are quite successful in reproducing the observed flux distribution of AGN. Similarly, the spatial distribution 
of polarization fraction appears to be dominated by the polar components of the model (Fig.~\ref{Fig:Maps}, top-right).
Scattering of disk emission by polar material naturally produces high polarization degrees, thus explaining the 
10 -- 20\% polarization within the winds. The northern part of the ejection flow (which is inclined towards the 
observer) has a polarization distribution that is not homogeneous due to projection effects. The central part 
of the winds is more strongly polarized than the edges, where the volumetric amount of matter is significantly 
smaller than along the polar axis, resulting in less light-matter interaction. Lower scattering probability and 
host depolarization leads to smaller observed polarization degrees, but a particular contrast is visible 
along the edges of the polar flow. Due to the de~Vaucouleurs flux distribution, dilution is less important 
at larger distances and the finite size of the outflow naturally shows higher $P$. This is in perfect agreement
with the observed distribution of polarization fraction in NGC~1068 \citep{Gratadour2015}. The polarized flux map 
(Fig.~\ref{Fig:Maps}, bottom-left) shaves off the stellar fluxes and highlights the AGN regions that present 
the highest amounts of flux and/or polarization degree. The central part of the image, where the supermassive 
black hole resides, is among the dominant sources of polarized fluxes but this is not due to a direct view 
towards the accretion disk. It is a combination effect of 1) high polarization fractions due to scattering 
of disk emission by the bases of the polar cones, 2) polarization from starlight by host dust grains, and 
3) strong stellar fluxes originating from the inner radius of the host extending up to the AGN torus. The 
cumulated effects of high fluxes and high polarization clearly highlight the central region. The edges of the 
polar winds also show high polarized fluxes, which is a border effect. Finally the map of the polarization 
position angles (Fig.~\ref{Fig:Maps}, bottom-right) shows a centro-symmetric component that is clearly visible 
on the edges of the lower wind cone (the upper cone showing the same signatures at larger distances, where 
dilution by the host becomes less important). This is a clear signature of scattered emission from the central 
engine. This pattern was already observed by \citet{Gratadour2015} and confirms the assumptions of the authors 
about the origin of this structure. The direction of the polarization angle pinpoints the source of emission 
and allows us to retrieve the location of the hidden nucleus \citep{Axon1996,Capetti1997,Kishimoto1999}. 

Our numerical polarimetric maps of NGC~1068 have shown very strong agreement with the observed flux and polarization 
distribution. The addition of a starlight component allows us to reproduce the expected amount of polarization 
and helps us to create realistic predictions for future observations. The possibility to extract polarimetric 
information of individual photons is the key to interpret the data. {\sc stokes} is able to decompose the 
results in order to better understand the contribution of each model component to the net polarization. Hence, 
by using the newest version of the code, it is possible to fit and interpret high-angular resolution polarimetric
data in a reliable way.

\section{Discussion}
\label{Discussion}

\subsection{Deriving the intrinsic polarization of type-2 AGN}
\label{Discussion:type2}
The new features we included in our code grant large possibilities to interpret data and fit observations. In 
particular, \citet{Kay1994} have shown that interpreting type-2 polarization is problematic due to the strong 
diluting effects of the host. Yet, using {\sc stokes}, it is now possible to retrieve the real amount of 
scattered light from the AGN only. For this purpose, the same simulation must be ran in two modes: with and without 
host dilution. Since the degree of polarization is a geometric value, it is independent on the photon flux
and thus a direct comparison between the polarization properties of the two models allows us to estimate the 
intrinsic polarization from the AGN. To compare the photometric and spectroscopic results, it is necessary 
to parametrize the bolometric fraction emitted by the different components. This is implemented 
by specifying the number of radiation sources and their relative strengths. A different technique to evaluate 
the true polarimetric signal of edge-on AGN was used in \citet{Marin2014b} and consists of dividing the
polarized flux of the AGN by the total flux across the broad emission lines. Only the broad lines polarization is a 
reliable indicator of the polarization of the scattered component but broad lines are, by definition, not 
detected in the total flux spectra of type-2 so proxies had to be used. With {\sc stokes} it is now possible
to overcome this problem. Polarized spectra are preferable but narrow-band polarimetric informations can also 
be used. We stress that degeneracies are still present, but coupling total flux, polarization degree and 
polarization position angle information allows one to better constrain the unknown AGN parameters than with spectroscopy
only. The addition of timing information is also crucial to remove further degeneracies \citep{Gaskell2012,Rojas2018}.
To narrow the uncertainties, observations of the host galaxy are needed in order to derive the fraction of
starlight to the total flux. \citet{Miller1983} have proven that it is possible to estimate this fraction by 
varying the ratio of the AGN and host spectra in order to cancel the absorption lines from stars. SED 
reconstruction from archival data is another alternative way but correction of the observed fluxes for 
aperture-dependence is crucial (Marin, submitted.).

\subsection{The AGN-starburst connection}
\label{Discussion:Starburst}
The numerical implementation of a galactic structure around the AGN has additional beneficial effects. It was 
shown in Sect.~\ref{Modeling:maps} that the host galaxy can be modeled with various geometries and compositions,
which allows us to implement radial distributions of emitting sources. The long-standing problem of the interplay 
between star formation and AGN activity may thus be explored. It is well known that there is a tight relationship
between the accretion activity of supermassive black holes and the evolution of their host galaxies \citep{Cattaneo1999,Ferrarese2000,Tremaine2002}
but it is often difficult to separate the contribution of the AGN from starburst light. The mixing between the two 
can be probed using emission-line ratio diagnostics. In particular, \citet{Davies2014a,Davies2014b} found 
a smooth transition from AGN-dominated emission in the center of the host to pure star-forming activity at 
larger galacto-centric distances. The relative contributions of star formation and AGN activity across AGN
host galaxies tells us about the evolutionary mechanisms of galaxies and allow us to probe the condition of 
the interstellar medium over cosmic time \citep{Davies2017}. Optical and near-infrared polarimetric observations 
of starburst-dominated AGN are less common, but \citet{Brindle1991} undertook a study of 27 starburst and interacting 
galaxies. Among other results, they found that their polarization is very low, i.e. $\le$~1\%, and decreases 
with increasing wavelength. The most likely physical mechanism to produce such low, wavelength-dependent 
polarization degrees is the transmission of light through aligned dust grains. The dichroic explanation is 
also favored by \citet{Draper1994} to explain the observed the optical polarization of NGC~1808. The polarization 
degree is found to decrease with increasing radial distances from the center of NGC~1808 and, similarly to the 
radial-dependent emission-line ratio diagnostics of \citet{Davies2014a}, a radial-dependent polarimetric 
investigation can be undertaken with {\sc stokes} in the future. It will be possible to test whether  
Low Ionization Nuclear Emission Region (LINER) Ultraluminous infrared galaxies (ULIRG) are necessarily 
starburst-dominated \citep{Antonucci2002}.

\subsection{Extension towards the infrared}
\label{Discussion:IR}
A critical limit of the code is that it is only valid up to 1~$\mu$m, where dust thermal re-emission starts
\citep{Sanders1989,Wilkes2004}. This hampers any investigation of the infrared polarization of AGN. The 
infrared band holds several key information on the physics of AGN, the host galaxy's nuclear environments, 
and the evolution of the system through time. In particular it is the best waveband to test the alignment of 
dust grains by parsec-scale magnetic fields. \citet{Lopez2015} measured the J' and K' polarization of NGC~1068 
and found that the large infrared polarization (7.0 $\pm$ 2.2\%) is likely due to a 4 -- 150~mG magnetic field,
depending on several physical assumptions. To probe the galactic magnetic field direction and strength, 
infrared polarimetric modeling is thus mandatory. {\sc stokes} was used as a comparison point between 
several radiative transfer codes that work in the near and mid-infrared, such as {\sc skirt} \citep{Peest2017}
or MontAGN \citep{Grosset2016,Marin2017,Grosset2018}. The results are in good agreement around 1~$\mu$m
and it is possible to couple those codes with {\sc stokes} for a first-order estimation of the ultraviolet-to-infrared
polarization of a given AGN model. However no other code than {\sc stokes} is able to handle such detailed 
physics, nor our newest implementations (starlight and ISP). Hence the development of {\sc stokes} towards the 
infrared is among the priorities for the next updates.

\section{Conclusions}
\label{Conclusions}

We present a new version of {\sc stokes}, a Monte Carlo code that allows one to simulate the radiative transfer of 
photons in a complex, three-dimensional environment. The main features we included to the code are the real yet 
parasitic components that dilute the observed signal of AGN : interstellar polarization and starlight photons from 
the host galaxy. It is well known that AGN polarimetric observations suffer from those extra components that can 
alter or even overshadow the true scattered AGN radiation. Our code now handles these two aspects in order to 
allow for fast and easy interpretation of observed results, together with a better understanding of the physical 
processes occurring at small scales. We have applied our code to NGC~1068, an archetypal type-2 AGN whose 
polarimetric signal is dominated by the host galaxy. We started by successfully reproducing the SED of the system 
at several scales (from nuclear fluxes to large apertures). Then, we used a generic model to emulate the 
spectropolarimetric signal of the AGN. Compared to archival data our code is now able to reproduce the 
wavelength-dependent total flux and polarized flux, together with the correct polarization angle. We also 
determined a putative inclination angle of the core engine that matches previous estimations. Finally, we used 
{\sc stokes} to reproduce the observed high-angular resolution polarization maps of NGC~1068. Our maps show 
a very detailed agreement with observations, replicating the generic flux and polarization levels together with small 
scales polarimetric features. Until our new code version, such fitting was not possible and simple first-order
corrections had to be done to estimate the diluting impact of the host galaxy.

In conclusion, the new version of {\sc stokes}, that will shortly be available through its dedicated website 
(http://www.stokes-program.info/), will help observers to better estimate the expected polarization from a 
source prior to any observation. Observational results will be easily compared to simulations that now account
for multi-scale dilution by the host. By running grids of models it will be possible to constrain several
unknown AGN parameters such as the system inclination, the size and geometry of the reprocessing regions, or 
the true scattered polarization from the inner components. We also highlight the fact that our code is not 
restricted to AGN, but can be applied to any cosmic sources that suffers from host and interstellar contamination.

\acknowledgements 
The author would like to thank Ren\'e W. Goosmann for his comments on the text, Ariane Lan\c{c}on for her 
advises on galaxy structures and physics, and Mari Kolehmainen for proof-reading this paper. This 
work was supported by the Centre national d'\'etudes spatiales (CNES) who funded this project through to 
the post-doctoral grant ``Probing the geometry and physics of active galactic nuclei with ultraviolet 
and X-ray polarized radiative transfer''.

\bibliographystyle{aa}
\bibliography{biblio}

\end{document}